\documentclass[aip,rsi,amsmath,amssymb,reprint]{revtex4-1}
\usepackage{graphicx}
\usepackage{dcolumn}
\usepackage{bm}
\usepackage[utf8]{inputenc}
\usepackage[T1]{fontenc}
\usepackage{mathptmx}
\usepackage{siunitx}

\DeclareSIUnit{\torr}{Torr}

\begin{document}
\thispagestyle{empty}
\pagenumbering{roman}
\onecolumngrid
\vspace*{5cm}
\noindent This article may be downloaded for personal use only. Any other use requires prior permission of the author and AIP Publishing. This article appeared in S.M.~Fleischer \textit{et al.}, \textit{Review of Scientific Instruments} \textbf{93}, 064505 (2022) and may be found at  https://doi.org/10.1063/5.0089933.
\clearpage\twocolumngrid
\pagenumbering{arabic}

\title{A Cryogenic Torsion Balance Using a Liquid-Cryogen Free, Ultra-Low Vibration Cryostat}

\author{S.~M.~Fleischer}%
\email{fleiscs3@wwu.edu}
\affiliation{Department of Physics and Astronomy, Western Washington University, Bellingham, Washington 98225, USA}
\author{M.~P.~Ross}
\author{K.~Venkateswara}
\author{C.~A.~Hagedorn}
\author{E.~A.~Shaw}
\author{E.~Swanson}
\author{B.~R.~Heckel}
\author{J.~H.~Gundlach}
\affiliation{Center for Experimental Nuclear Physics and Astrophysics, University of Washington, Seattle, Washington 98195, USA}

\date{\today}

\begin{abstract}
We describe a liquid-cryogen free cryostat with ultra-low vibration levels which allows for continuous operation of a torsion balance at cryogenic temperatures. The apparatus uses a commercially available two-stage pulse-tube cooler and passive vibration isolation. The torsion balance exhibits torque noise levels lower than room temperature thermal noise by a factor of about four in the frequency range of \SIrange{3}{10}{mHz}, limited by residual seismic motion and by radiative heating of the pendulum body. In addition to lowering thermal noise below room-temperature limits, the low-temperature environment enables novel torsion balance experiments. Currently, the maximum duration of a continuous measurement run is limited by accumulation of cryogenic surface contamination on the optical elements inside the cryostat.
\end{abstract}
\maketitle

\section{Introduction}

Starting with the original Cavendish experiment\cite{cavendish}, torsion balances have a long history as sensitive probes of weak forces. For modern torsion balance experiments using state-of-the-art technology and carefully controlled systematics, thermal noise represents one of the fundamental limitations\cite{adelberger:2009, wagner:2012}. In the case where damping is dominated by the intrinsic mechanical loss in the material of the suspension fiber, the loss term in the equation of motion for the torsion pendulum is velocity-independent and can be described by assuming a complex spring constant. Application of the fluctuation-dissipation theorem then yields an expression for the thermal torque noise\cite{thermal} as a function of angular frequency $\omega$:

\begin{equation}
    \label{eq:thermalnoise}
    S_\tau(\omega)=\sqrt{\frac{4 k_B T \kappa}{Q \omega}}
\end{equation}
where $k_B$ is the Boltzman constant, $T$ is the temperature, $\kappa$ is the torsional spring constant, and $Q$ is the quality factor of the torsion oscillation determined by the mechanical loss of the fiber. This noise level represents a fundamental limit to the performance of torsion balances and other precision experiments employing mechanical oscillators.

An examination of Equation~\ref{eq:thermalnoise} suggests three possible approaches to decreasing the thermal noise of the system: minimizing the strength of the torsional spring (small $\kappa$), choosing fiber materials with low mechanical loss, and lowering the temperature. Typical torsion balance experiments have already optimized the spring constant, leaving low-temperature operation and minimized mechanical loss as avenues for improved sensitivity. Previous research has shown that moving to cryogenic temperatures leads to a decrease in noise not only by lowering $T$ but also by significantly improving the quality factor of conductive metal fibers\cite{bantel:2000, newman:2014}. As another approach to minimizing mechanical loss, torsion fibers made from fused silica or crystalline quartz are very strong and can reach much higher $Q$ values at room temperature, which do not improve on cooling to cryogenic temperatures\cite{schroeter:2007}. Values of up to $\sim 10^8$ have been observed for bulk samples of fused silica at room temperature\cite{ageev:2004}. However, as electrical insulators, such fibers have a major disadvantage: they make it much harder to ensure that the pendulum body and its surroundings are held at the same electrostatic potential. This includes concerns about possible charge buildup on the pendulum body over long times. For these reasons we limit ourselves here to metallic fibers.

Precision laboratory tests of gravity and weak-force measurements are not the only areas of physics research with a strong interest in pushing the limits of thermal noise and mechanical vibrations. Several current design studies for future ground-based gravitational wave detectors like LIGO Voyager\cite{voyager} and the Einstein Telescope\cite{et} include plans for cooling the test masses to cryogenic temperatures, with extremely severe constraints on allowable mechanical vibration levels\cite{ligo_whitepaper2020}.

Here we describe an apparatus which allows for the operation of a torsion balance at cryogenic temperatures with minimal maintenance, using only passive vibration isolation and without the need for liquid cryogens. Its ultra-low vibration levels make it possible to exploit the reduction in thermal noise to perform sensitive mechanical experiments.

\section{Apparatus}
\subsection{Overall Design\label{sec:overall-design}}

The main challenge in designing the cryostat lies in a torsion balance's sensitivity as an instrument to measure extremely weak forces. Even though the torsional mode is relatively well decoupled from the more easily excitable swing and bounce modes, great care needs to be taken to avoid even low levels of mechanical disturbances. In order to achieve the goal of improving noise performance, the introduction of mechanical vibrations into the system has to be strenuously avoided. Unfortunately, any closed-cycle cryocooler is a significant source of vibrations. Pulse-tube coolers have been reported to be more quiet than Gifford-McMahon coolers: a measurement by Wang and Gifford \cite{wang_gifford} showed a reduction of the forces measured at the mounting flange by a factor of 50. Furthermore, pulse-tube coolers with a remote rotary valve unit are available, which further reduces vibrations. Nevertheless, due to the expansion and contraction of the tubing with every helium pulse, their cold flange still performs a cyclic motion with an amplitude of the order of microns\cite{suzuki:2006}. Novel ideas for reducing the level of vibrations have been proposed, for example by combining two out-of-phase pulse-tube coolers connected to a single cold flange\cite{suzuki:2006}. However, such systems are not readily available. We designed our cryostat to use a standard two-stage pulse-tube cooler providing \SI{0.5}{W} of cooling power at \SI{4}{K} (Sumitomo RP-062B)\cite{shi}. We chose an approach inspired by a cryostat described by Caparrelli \emph{et al.}\cite{caparrelli:2006}, but we opted for using only passive vibrational isolation, with a design which could accommodate active isolation if necessary.

\begin{figure}[t]
\centering \includegraphics[width=0.48\textwidth]{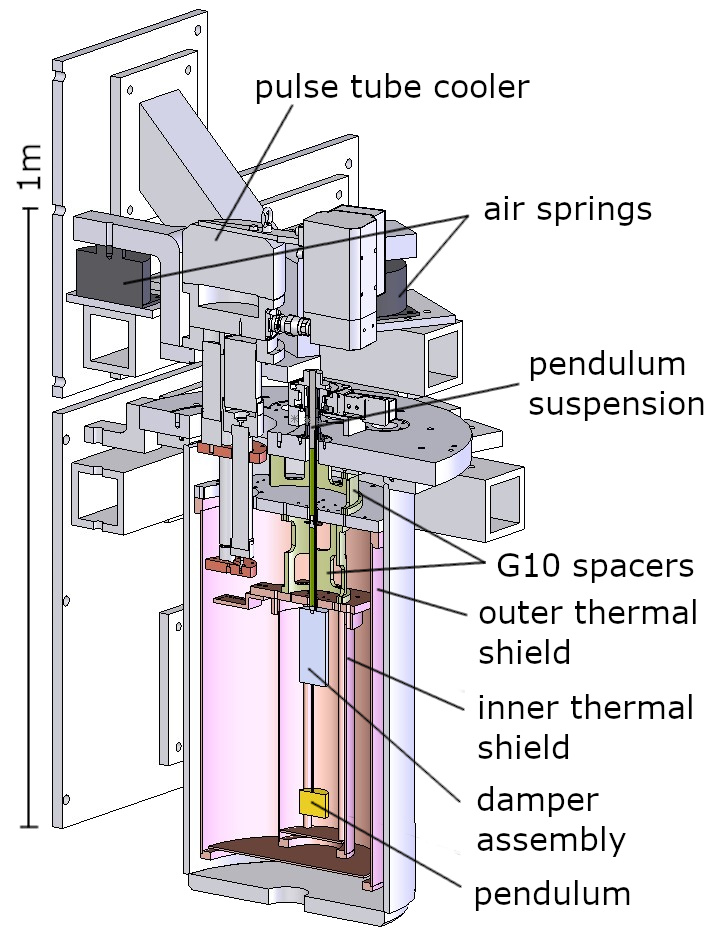}
\caption{Cutaway view of the overall design of the experimental apparatus. Not shown are: the heat links, the magnetic shield, the bellows connecting cold head and vacuum chamber, the autocollimator, wiring and instrumentation, and fasteners and various other small parts. The pendulum, fiber, damper, and positioning system are only shown schematically.}
\label{fig:apparatus} 
\end{figure}

Figure~\ref{fig:apparatus} shows a cutaway view of the main components of the apparatus. The cryostat is rigidly mounted to the massive, well-settled concrete wall of a former cyclotron cave in a partially underground lab at the Center for Experimental Nuclear Physics and Astrophysics (CENPA) at the University of Washington. It is held by two separate supports, one for the main vacuum chamber of the cryostat containing the torsion balance and another one above it for the pulse-tube cooler. The separation of the support structures helps to minimize the vibrational coupling between the pulse-tube cooler and the cryostat. Both support structures are significantly oversized for their loads to provide maximum stiffness and to avoid any low-frequency resonant modes. The supports are securely bonded to the wall using epoxy mortar and steel bolts. The main body of the cryostat is rigidly mounted on its support via three adjustable legs. The cold head is resting on three air springs (GMT Gummi-Metall-Technik GmbH, model LF13010890) mounted atop its separate support structure. The air springs provide a means of achieving a low resonance frequency of \SI{\sim 3}{Hz} while maintaining a compact form factor. The cryostat body and the cold head are connected only by flexible heat links and an edge-welded bellows seal at the top flange. This arrangement --- a rigidly mounted cryostat with a ``floating'' cold head and as little coupling between the two as possible --- was chosen to allow for precise leveling of the torsion balance setup and to guard against problematic long-term drifts. As it is a strong source of mechanical vibrations, the helium compressor for the pulse-tube cooler is located in an adjacent room, about 5 meters away from the apparatus.

The main body of the cryostat is a cylindrical vacuum chamber suspended from its top flange. It is held at \SI{\sim 0.1}{\micro\torr} via continuous pumping with a turbo pump and a dry roughing pump; during low-temperature operation, the large surface area of the thermal shields (see Sec.~\ref{sec:cryo} below) provides additional cryo-pumping. For this reason and due to the convoluted nested structure of the cryostat described below, the residual gas pressure inside the cold volume is likely lower than measured on the outside during low-temperature operation. On the other hand, it is to be expected that this same nested structure leads to a substantially worse vacuum in the payload volume when used at room temperature. To minimize vibrational coupling, the turbo pump is mounted on a separate stand on the floor instead of directly on the vacuum vessel. Its connection to the vacuum chamber consists of two flexible stainless steel bellows and a straight section which is clamped between two bored-out lead bricks bolted to the wall for vibration damping.

Immediately inside the vacuum vessel, separated by a gap of a few millimeters, a cylindrical magnetic shield made from perfection-annealed Co-NETIC AA\cite{conetic} fully surrounds the inside volume of the cryostat, only featuring a number of circular holes in its top and bottom surfaces for feedthrough penetrations and pumping. Two loops of Kapton-insulated copper wire threaded through the magnetic shield and connected to an electrical current feedthrough allow for degaussing of the shield. Additionally, the cylindrical surface of the inner thermal shield (see Section~\ref{sec:cryo}) is wrapped in lead foil. During low-temperature operation, the inner shield temperature drops below the \SI{7.2}{K} critical temperature for superconductivity in lead, providing a shield against time-varying magnetic fields.

All necessary power and instrumentation feedthroughs are located on the top plate of the vacuum chamber. Together with the design of the thermal shields described below, this allows for the cylindrical vacuum vessel and all shields to be easily removable when accessing the inside of the cryostat. The torsion balance itself is enclosed completely within the cold volume of the inner thermal shield. A viewport provides optical access to the torsion pendulum for an autocollimator to monitor its angular position.

\subsection{Thermal Considerations\label{sec:cryo}}

\begin{figure}[t]
\centering \includegraphics[width=0.48\textwidth]{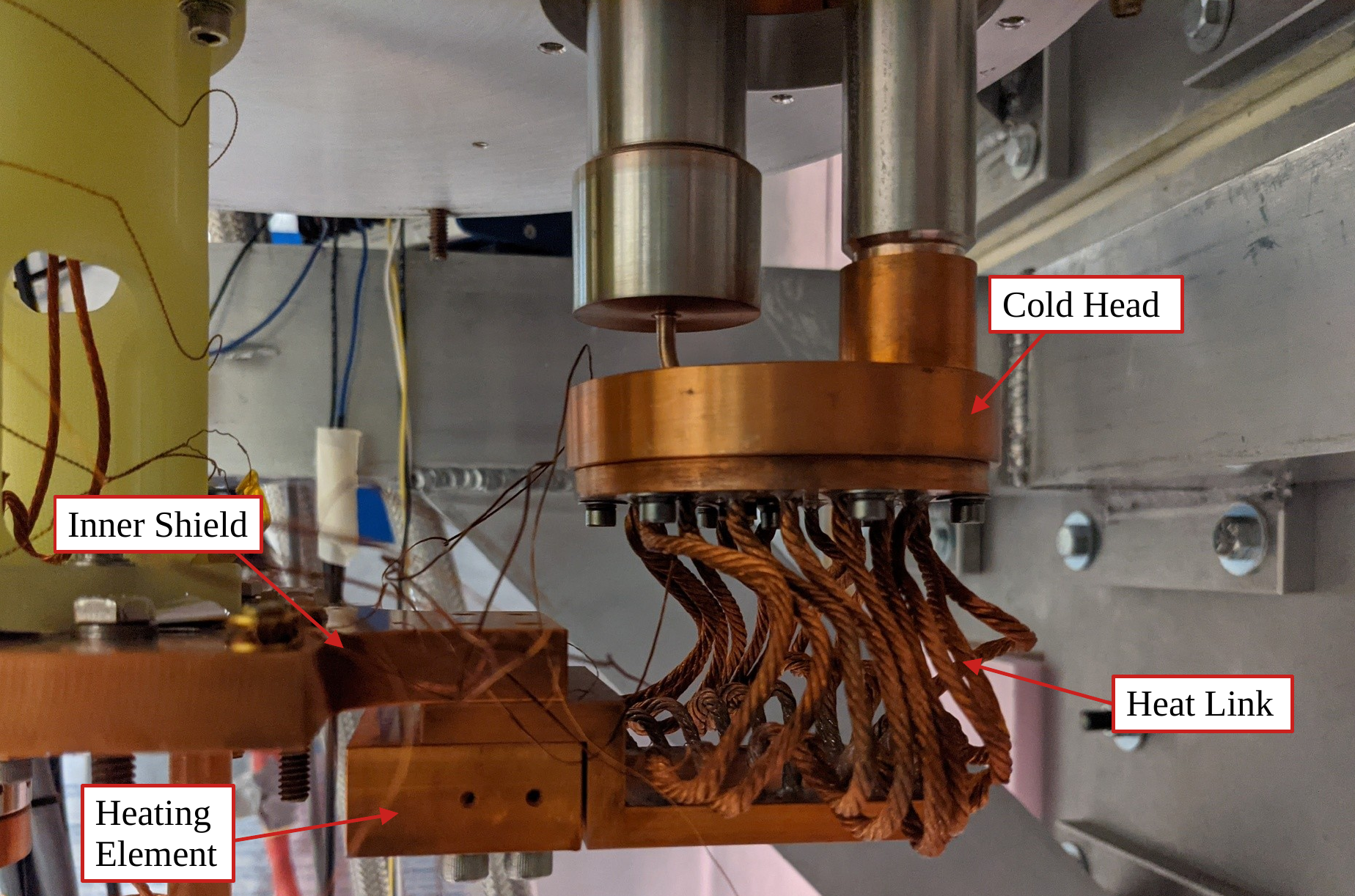}
\caption{A photograph showing the flexible heat link connecting the inner thermal shield to the second stage of the cold head. The top plate of the thermal shield is partially visible in the lower left corner (the cylindrical part of the shield has been removed). The stranded copper conductors of the heat link can be seen near the center of the frame. As described in the text, they are soft-soldered to copper flanges on either end, which are then bolted to the cold head and to the thermal shield, respectively. The rectangular copper block located at the latter connection contains a heating element which can be used for testing purposes or to operate the instrument at temperatures above the minimum of \SI{\sim 5}{K}.}

\label{fig:heatlink} 
\end{figure}

The two-stage pulse-tube cooler is specified to provide \SI{0.5}{W} of cooling at \SI{4}{K} (stage 2) and \SI{30}{W} at \SI{50}{K} (stage 1) with ``turn-key'' operation and very little maintenance. Based on these specifications and the geometrical constraints posed by the cold head, the cryostat features a fairly typical design. It consists of an aluminum outer vacuum chamber containing two nested copper thermal shields thermally coupled to the two stages of the cold head. The top plates of the vacuum vessel and of the thermal shields are fixed in place, both of the latter supported by G10 spacers hanging from the respective previous stage (see Fig.~\ref{fig:apparatus}). To minimize vibrational coupling, thermal contact between the shields and the cold head is made using flexible heat links. They consist of lengths of flexible stranded copper conductors (ABL Electronics Supplies, Inc. Extra Flexible Bare Copper Rope Lay, AWG 10, part no. 97010), made from 413 AWG 36 wires each. On either end, the copper wires are soft-soldered to blind holes in a copper flange. Care was taken to avoid compromising the flexibility of the stranded copper by solder wicking along their length. The best results were achieved by carefully heating the flange with all wires inserted on an electric hot plate and then applying rosin-core flux lead-tin solder to each hole. Figure~\ref{fig:heatlink} shows the heat link connecting the cold head flange to the inner shield. An electric heating element potted into a copper block with Stycast 2850FT is mounted at the warm side of the inner heat link. It allows for operation of the instrument at temperatures above the minimum, or it can be used for diagnostic purposes. For the data presented in this paper, the heater was not used. Several layers of super-insulation wrapped around the outer thermal shield reduce the radiative heat load on the outer thermal shield. Minimizing the heat load is even more important for the inner shield; due to the much lower heat capacity at low temperatures and due to the limited cooling power available from the second stage of the pulse-tube cooler, leakage of room-temperature thermal radiation has to be avoided. While the shields require openings to the remainder of the vacuum chamber for pumping, care was taken to avoid any direct line-of-sight penetrations through them. For example, the bottom of each shield is a disk slightly smaller in diameter than the inner diameter of the shield. The resulting gap is overlapped by an annular radiation baffle soldered to the cylindrical part of the shield, leaving a few millimeters of vertical space to the bottom disk.

\begin{figure}[t]
\centering \includegraphics[width=0.48\textwidth]{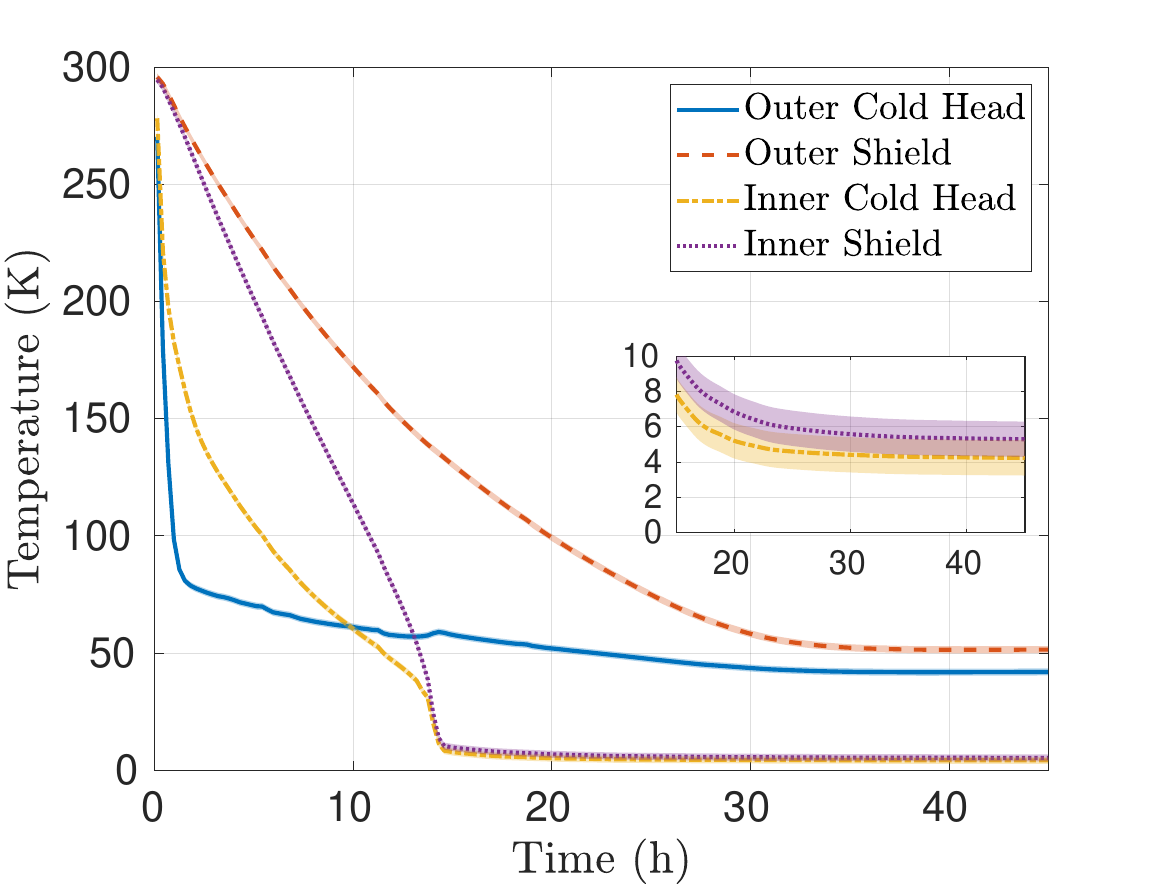}
\caption{Temperatures of the first and second stage of the cold head as well as of the outer and inner thermal shield of the cryostat during a cool-down of the apparatus from room temperature to low-temperature equilibrium. The inset shows an enlarged view of the region around the final inner shield temperature. The shaded bands indicate the temperature sensor accuracy of \SI{\pm 1}{K}.}
\label{cooldown} 
\end{figure}

For temperature monitoring, several Si-diode temperature sensors (LakeShore Cryotronics DT-670) are used in conjunction with quad-lead phosphor-bronze wiring thermally anchored to the thermal shields. After starting the cryocooler, the apparatus reaches stable operating temperature within about two days. Figure \ref{cooldown} shows the temperatures of both stages of the cold head and of both thermal shields as a function of time during a cool-down to cryogenic temperatures. The usable low-temperature volume is a cylinder of \SI{127}{mm} diameter and \SI{375}{mm} height. It reaches an equilibrium temperature of \SI{\sim 5}{K}. 

The suspension point of the torsion balance requires motorized positioning capability with a few centimeters of linear vertical travel as well as about \ang{\pm 180} of rotation around a vertical axis. This mechanical access to the cold volume is achieved using regular in-air, room-temperature positioning stages on top of the vacuum vessel with a rigid connection reaching into the inner shield. The connection is made using a \SI{12.7}{mm}-diameter G10 tube with thermal breaks located at the penetrations through the thermal shields. The thermal breaks consist of copper disks with a large enough diameter to avoid creating direct line-of-sight openings into the shields, heat-sunk to the thermal shields using flexible stranded copper wire (ABL Electronics Supplies, Inc. Extra Flexible Bare Copper Rope Lay, AWG 14, part no. 97014), soldered in the same manner as described earlier with sufficient slack to allow for the necessary amount of movement. 

As the torsion pendulum is suspended from a thin fiber, cooling of the pendulum body takes places only radiatively and through the limited heat conduction along the fiber. For efficient heat removal during cool-down, the pendulum can be lowered onto a pedestal at the bottom of the inner thermal shield (the ``parking stop''). For some future experiments it may be critically important to reach and maintain a pendulum temperature below a certain threshold, for example below the critical temperature of a superconductor such as niobium. Our tests have shown that this can be ensured by introducing \SI{\sim 1}{\micro Torr} of helium into the vacuum chamber to allow for convective cooling. For the measurements discussed in this paper, no helium exchange gas was used; all data refer to the high-vacuum configuration as described above. While it would be very interesting to measure the temperature of the pendulum body of the torsion balance directly, we were unable to do so without disrupting its function and significantly altering its equilibrium temperature.

To prevent room temperature fluctuations from influencing the pendulum suspension or the leveling of the vacuum chamber via differential thermal expansion of the various materials, the cryostat is surrounded by a temperature-stabilized enclosure. A feedback loop stabilizes the air temperature inside it, reducing the measured temperature fluctuations of the top plate of the vacuum vessel to approximately \SI{\pm 0.01}{K}. This is achieved using a recirculating temperature controller in the adjacent room pumping water through two heat-exchanger/fan units inside the enclosure. The vacuum pumps and the electronics rack are placed outside of this enclosure.

\begin{figure}[t]
\centering \includegraphics[width=0.45\textwidth]{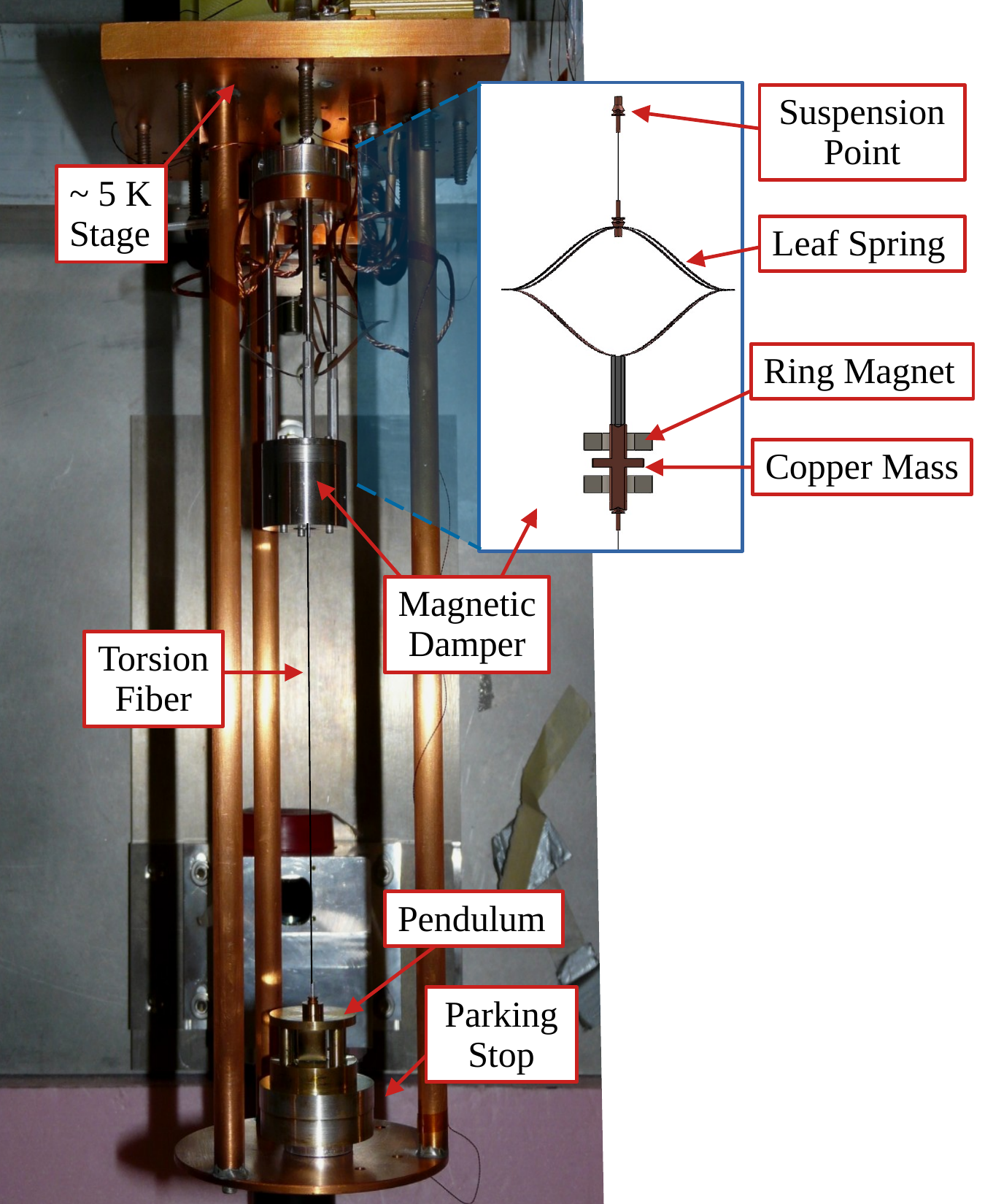}
\caption{A photograph of the torsion balance with the vacuum vessel, the magnetic shield, and both thermal shields removed. The torsion balance consists of the pendulum body suspended from a \SI{30}{\micro m}-diameter torsion fiber (highlighted for clarity), which in turn is attached to the magnetic damper. The damper is mounted to the mechanical positioning system described in Section~\ref{sec:cryo} and thermally anchored to the top of the innermost thermal shield, held at roughly \SI{5}{K}. In the photo, the pendulum is temporarily set down onto the cylindrical parking stop. The inset shows a schematic drawing of the magnetic damper assembly, which employs a small copper mass suspended in a cylindrically symmetric magnetic field to damp the swing and ``bounce'' degrees of freedom while maintaining a high $Q$ for the torsion mode.}

\label{fig:pendulum} 
\end{figure}

\subsection{The Torsion Balance}
The torsion balance, shown in Figure \ref{fig:pendulum}, is located inside the cold volume of the inner shield of the cryostat. It consists of a pendulum body suspended from a \SI{165}{mm}-long, \SI{30}{\micro m}-diameter beryllium-copper fiber (California Fine Wire Company Beryllium Copper Alloy 25 CDA 172, operating at $\sim 80\%$ of its yield strength) attached to a magnetic swing damper. The connections to the torsion fiber on both sides are made using copper crimp tubes (outer diameter \SI{0.71}{mm}, inner diameter \SI{0.10}{mm}, Saturn Industries). The room-temperature positioning system described in Section~\ref{sec:cryo} controls the position of the damper, which serves two functions: it provides passive seismic isolation, and it decreases the amplitude of the swing mode while maintaining a high quality factor for the torsional oscillation. The damper employs a small copper mass suspended inside a cylindrically symmetric magnetic field generated by two samarium-cobalt permanent ring magnets. The suspension of the copper mass consists of a set of beryllium-copper leaf springs and a short tungsten fiber. With this setup, any unwanted vertical or horizontal motion stemming from the swing or ``bounce'' degrees of freedom drives eddy currents and produces a damping force. Due to the symmetrical design, the torsional oscillation is largely unaffected by the magnetic damper. To avoid stray magnetic fields influencing the torsion pendulum, the permanent magnets are surrounded by a cylindrical iron housing acting as a flux return. A system very similar to the one employed here has been described in detail elsewhere\cite{MINT}.

A heater tape wrapped around the outside of the vacuum vessel of the cryostat allows for a mild bake-out of the complete system to about \SI{40}{\celsius}. This helps to minimize drift in the pendulum's equilibrium position by speeding up initial relaxation processes which lead to a slow ``unwinding'' of the fiber whenever the pendulum is first suspended.

The pendulum, shown in Figure~\ref{fig:pendulum}, has a mass of \SI{51}{g} and a moment of inertia of \SI{1.13e-5}{kg\cdot m^2}. Special attention was given to minimizing unwanted interactions between the pendulum and the environment. In particular, the pendulum was designed to minimize its relevant low-order mass multipole moments\cite{multipoles} to avoid gravitational coupling to fluctuating gravity gradients, such as those caused by people nearby.

With the parameters given, the resonance frequency of the free torsional oscillation is \SI{7.05}{mHz}, with observed quality factors $Q$ varying between $\num{3.2e4}$ and $\num{4.7e4}$, depending on the exact fiber used. This corresponds to a torsional spring constant of $\kappa=\SI{2.22e-8}{N m/rad}$ and an amplitude decay time of \SIrange{16.7}{24.6}{days}, respectively. For comparison, we typically find quality factors of a few thousand for metallic torsion fibers at room temperature. Notably, the magnetic shield described in Sec.~\ref{sec:overall-design} is required to achieve these $Q$ values. Without such a shield, eddy current damping due to the Earth's magnetic field limits the observed quality factor to much lower values.

\subsection{Optical Readout}

The angular position of the pendulum, which has a four-fold symmetry with mirrors on each side, is recorded using a multi-slit autocollimator \cite{MSA} attached to the concrete wall below the supports of the apparatus. This type of autocollimator has the ability to make differential angle measurements with respect to a fixed reference mirror. To allow for subtraction of common mode noise due to motion of the optics, the vacuum chamber, and the thermal shields, the reference is mounted inside the cryostat. Placements both on the inner and the outer thermal shield were tested, with no clear advantage in common noise rejection for either location. 

The optical angle readout requires a clear line of sight from the room-temperature environment directly to the cold pendulum body. To allow for this, both thermal shields have \SI{12.7}{mm}-diameter openings for the autocollimator light beam. Initially, we employed a simple system of radiation baffles to limit the size of the opening and the solid angle for room-temperature thermal radiation to enter the cold volume of the inner shield. In this configuration, heating due to room-temperature thermal radiation prevented cooling of the pendulum below temperatures of about \SI{100}{K}. This is a consequence of the fact that cooling of the pendulum body is possible only radiatively or via the very limited heat conductivity of the thin fiber. In order to maintain the necessary optical access while simultaneously reducing the heat load due to thermal radiation, a cold shortpass filter made from Schott KG2 glass was placed in the opening of the inner thermal shield.

\section{Results and Discussion}

\begin{figure}[t]
\centering \includegraphics[width=0.48\textwidth]{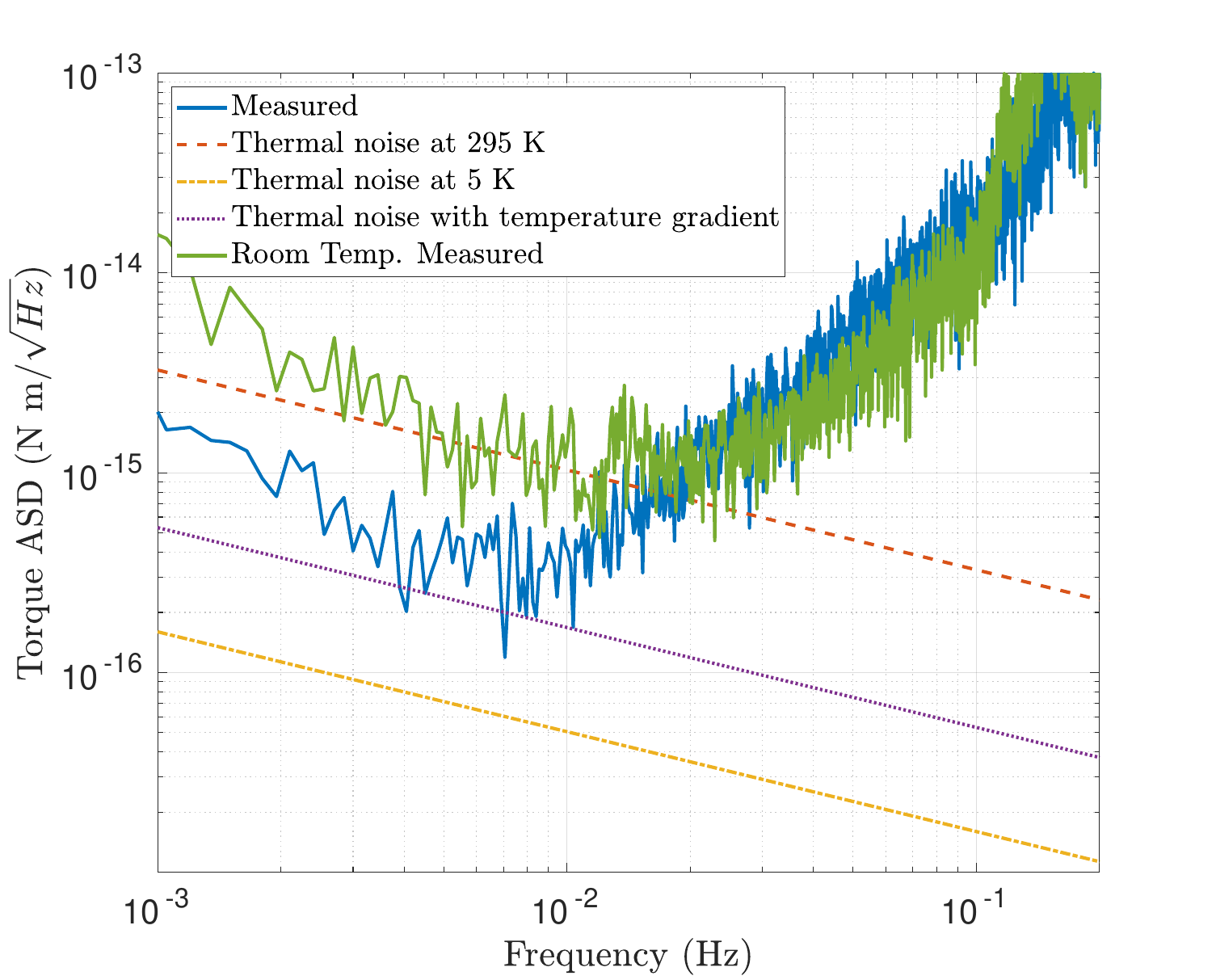}
\caption{Amplitude spectral density of the observed torque noise in the frequency range \SIrange{1}{100}{mHz}. The top and bottom straight lines show the theoretical thermal noise limits calculated from Equation~\ref{eq:thermalnoise} for room temperature (295 K) and for the temperature of the inner thermal shield (5 K). Corresponding to our measurements at these temperatures, the $Q$ values were assumed to be $6.0\times 10^3$ and $4.5\times 10^4$, respectively. Further, an estimated thermal noise level for a fiber with a temperature gradient from \SI{5}{K} at the suspension point to \SI{100}{K} at the pendulum body is shown (for details, see Appendix~\ref{app:noise}). The strong increase with frequency above \SI{10}{mHz} is due to readout noise.}

\label{fig:torque_noise} 
\end{figure}

\subsection{Noise Performance}

The multi-slit autocollimator achieves an angular sensitivity of \SI{\sim 3}{nrad/\sqrt{Hz}} above \SI{10}{mHz}. A spectrum of the torque noise of the apparatus is shown in Figure~\ref{fig:torque_noise}. Between \SI{3}{mHz} and \SI{10}{mHz} the noise is approximately a factor of four below room temperature thermal noise. Above \SI{10}{mHz} the torque noise strongly increases with frequency; in this region the system is limited by readout noise. 

While the observed noise spectrum shows a substantial improvement over room-temperature thermal noise, it clearly does not correspond to thermal noise at \SI{5}{K}, the temperature of the inner shield. Apart from the already mentioned readout noise, which dominates at  frequencies above \SI{10}{mHz}, we found a clearly detectable residual coupling to seismic motion. It is further worth noting that we were unable to directly measure the temperature of the pendulum body or the torsion fiber without significantly altering it. While changes in oscillation frequency, mostly due to the temperature-dependent modulus of the fiber, can serve as a proxy for the temperature of the fiber, this method is not well calibrated. Considering that residual room-temperature thermal radiation penetrating the shortpass filter and the autocollimator light beam impart a non-zero heat load on the pendulum, it appears safe to assume that its equilibrium temperature is somewhat higher than that of the inner shield. Figure~\ref{fig:torque_noise} also shows an estimated thermal noise limit for this case, assuming that the pendulum remains at a much higher temperature of about \SI{100}{K} (see Appendix~\ref{app:noise} for details). Considering these observations, the measured torque noise is likely a combination of residual seismic motion, the thermal noise at the fiber temperature, and readout noise. 

\subsection{Cryogenic Surface Contamination}

\begin{figure}[!h]
\centering \includegraphics[width=0.48\textwidth]{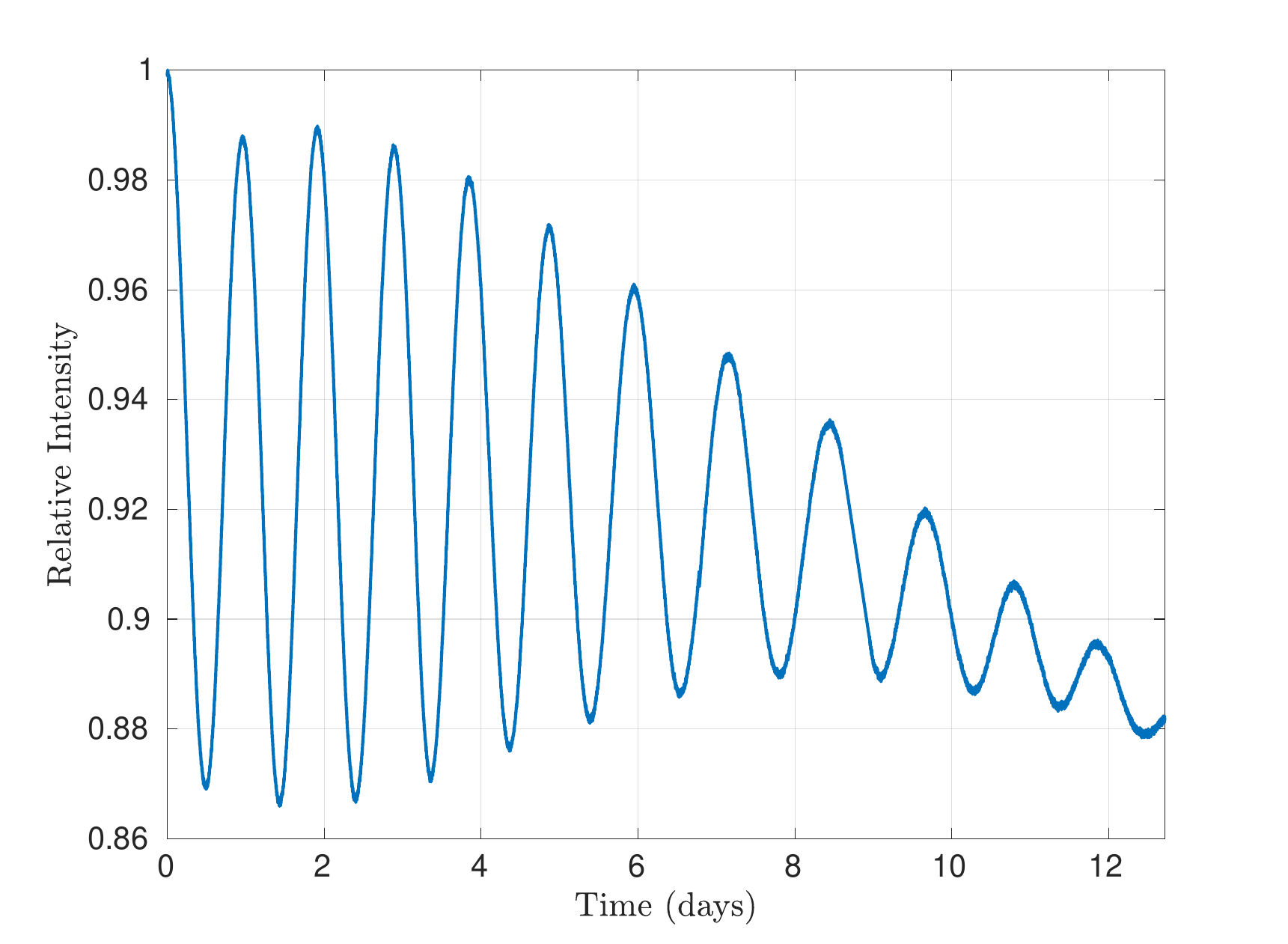}
\caption{Light intensity reflected by the reference mirror as a function of run time at low temperature. The intensity has been normalized to its value at the beginning of the run. The oscillating pattern with time is likely due to thin-film interference effects caused by an ice layer forming on a cold optical surface. The inferred buildup rate is \SIrange{0.22}{0.25}{\micro m/day}.}
\label{fig:ice} 
\end{figure}

Over extended periods of operation at low temperature a deterioration in performance of the optical readout is observed. Figure \ref{fig:ice} illustrates this problem by showing the time evolution of the light intensity reflected by the reference mirror over a period of 12 days. 

Our observations are compatible with a thin layer of residual gas freezing out on the optical elements inside the cryostat. As the thickness of this layer increases over time, the reflected light intensity exhibits an oscillating pattern of alternating interference fringes. At the same time, an overall decreasing trend indicates a reduction of the reflectance/transparency of the cold mirrors and/or the shortpass filter. A similar problem, although on a different scale, has been observed on cold optics for the cryogenic gravitational wave interferometer KAGRA\cite{hasegawa:2019}. As discussed in the cited reference, most likely the material deposited on the optical surfaces is residual water vapour present in the vacuum system or desorbing from the room-temperature surfaces. From the period of oscillation in the reflected reference beam intensity, the \SI{617}{nm} wavelength of the autocollimator light source, and the index of refraction reported by the KAGRA group\cite{hasegawa:2019} ($n=1.26$), we infer an ice layer buildup rate ranging from \SIrange{0.22}{0.25}{\micro m/day}. As seen in Fig.~\ref{fig:ice}, the slightly higher growth rate occurs at the beginning.

The observed deterioration of the optical readout signal limits the possible length of a measurement run to a few weeks. Full performance is recovered by thermally cycling the apparatus to room temperature and back to cryogenic temperature. While we have not yet made any systematic attempts to address this problem, it may be worth noting that various other mitigation strategies have been proposed in the context of cryogenic gravitational wave detectors\cite{spallino:2021}.

\section{Conclusions}

We have presented the design of an ultra-low vibration cryostat for high-precision mechanical experiments. The apparatus does not require liquid cryogens, offers very simple operation and can maintain cryogenic temperatures with minimal maintenance. The setup is used to operate a torsion balance, achieving torque noise levels about a factor of four lower than the room temperature thermal noise limit. Currently the noise performance of the torsion balance is limited by a combination of residual seismic noise, readout noise, and thermal noise due to the heat load on the pendulum. The maximum duration of a measurement run is limited by surface contamination accumulating on the cold parts of the readout optics.

In addition to the configuration described in this paper, many variations are possible in terms of torsion fiber materials, pendulum geometry, etc. Besides reducing thermal noise below room-temperature limits, the apparatus enables new torsion balance experiments such as tests of the equivalence principle for superconducting materials\cite{Tajmar} or searches for spin-dependent forces using a superconducting shield\cite{spin}.

\section{Acknowledgments}
The authors gratefully acknowledge financial support from the National Science Foundation, who funded this work under grant awards PHY-0653863, PHY-0969199, PHY-1305726, PHY-1607385, PHY-1607391, PHY-1912380, and PHY-1912514, and from the DOE, who funds the University of Washington Center for Experimental Nuclear Physics and Astrophysics. Additionally, S.F. would like to thank the Alexander von Humboldt Foundation for support as a Feodor Lynen Fellow. Many thanks are due to other members of the E\"ot-Wash group and to the staff of the CENPA and Physics Machine Shops at the University of Washington. Especially we would like to thank Hendrik Simons and Jim Elms for lending their great expertise and their skills to this project. David Hyde, Doug Will, and other CENPA staff deserve thanks for their help with various aspects of this work. We thank Massimo Bassan (University of Rome Tor Vergata) for his help while on sabbatical at the University of Washington and for his loan of a cryogenic accelerometer during early tests of the cryostat. 

\section{Data Availability}
The data that support the findings of this study are available from the corresponding author upon reasonable request.

\appendix
\section{Estimated Thermal Noise for a Fiber with a Temperature Gradient\label{app:noise}}
For a torsion balance with a uniform temperature along the suspension fiber, the amplitude spectral density of the thermal torque noise is given by Eq.~\ref{eq:thermalnoise}. If the pendulum body remains at a higher temperature due to a non-zero heat load in conjunction with the limited thermal conductivity of the fiber, there will be a temperature gradient along the fiber. Assuming a linear gradient from a temperature $T_0 = \SI{5}{K}$ at the suspension point to an unknown temperature $T_1$ at the pendulum body, we can estimate the expected torque noise.

Measurements by Duffy\cite{duffy:1992} show a temperature dependence of the quality factor for beryllium copper in the range from \SI{5}{K} to \SI{100}{K} which is well approximated by a power law. Fitting the data with an empirical function $Q(T) = (aT)^{-b} + c$ gives $a = \SI{4.7e-9}{K^{-1}}$, $b = 0.65$, and $c = 2.4\times 10^4$ as best fit parameters. We note that our observed room-temperature quality factor for the torsion oscillation is lower than expected from these data. On the other hand,
Bantel and Newman\cite{bantel:2000} report $Q$ values for a very similar torsion fiber as ours (same alloy and manufacturer, diameter \SI{20}{\micro m}) at \SI{4.2}{K} and \SI{77}{K}, which agree within a factor of 2 with Duffy's measurements, despite the very different geometry and surface-to-volume ratio. There is also a noticeable variation in $Q$ between different fibers prepared from the same wire stock. To achieve at least a rough estimate, we base our calculation in the following on the data published by Duffy.

As thermal noise power depends on $T/Q$, a noise estimate for a fiber with a temperature gradient requires integration over the fiber. If we use $x$ to denote the position along the torsion fiber of length $L$, we can write the temperature as a function of position as $T(x)=T_0+(T_1-T_0)x/L \equiv T_0 + \beta x$ and $Q$ as

\begin{eqnarray}
    Q(x) &=& Q(T(x)) = \left(a(T_0 + \beta x)\right)^{-b} + c \\
         &=& (aT_0)^{-b} \left[(1+\beta')^{-b} + c'\right]
\end{eqnarray}

with $\beta' \equiv \beta/T_0$ and $c' \equiv c(aT_0)^b$.

We can then use Eq.~\ref{eq:thermalnoise} and integrate over the length of the fiber to average the power spectral density of the torque noise:

\begin{eqnarray}
    \overline{S_\tau^2}(\omega) &=& \frac{1}{L}\frac{C}{\omega}\int_0^L \frac{T(x)\; dx}{Q(x)}\\
    &=& \frac{C}{\omega}\frac{(aT_0)^b}{L}\int_0^L\frac{T_0+\beta x}{(1+\beta'x)^{-b} + c'} dx\\
    &=& \frac{C}{\omega}\frac{(aT_0)^bT_0^2}{L\beta}\int_1^{1+\beta'L}\frac{u\;du}{u^{-b} + c'}
\end{eqnarray}

where $C = 4 k_B \kappa$, and in the last step we substituted $u = 1 + \beta'x$. The remaining integral can be written in terms of the Gaussian hypergeometric function:

\begin{equation}
    \int\frac{u\;du}{u^{-b} + c'} = \frac{u^2}{2c'}{}_2F_1\left(1, -\frac{2}{b}; \frac{b-2}{b};-\frac{u^{-b}}{c'}\right)
\end{equation}

In the same way, we can also calculate an averaged mechanical loss $\overline{Q^{-1}}$ for the fiber, giving
 
\begin{equation}
    \overline{Q^{-1}} = \frac{1}{L}\int_0^L \frac{dx}{Q(x)}
    = \frac{(aT_0)^bT_0}{L\beta}\int_1^{1+\beta'L}\frac{du}{u^{-b} + c'}
\end{equation}

where the integral can again be written in terms of the hypergeometric function: 

\begin{equation}
    \int\frac{du}{u^{-b} + c'} = \frac{u}{c'}{}_2F_1\left(1, -\frac{1}{b}; \frac{b-1}{b};-\frac{u^{-b}}{c'}\right).
\end{equation}

Evaluating these expressions requires an assumption about the unknown temperature of the pendulum body. If we simply assume that the pendulum body remains at a temperature of $T_1\approx \SI{100}{K}$, we obtain an averaged quality factor $\overline{Q}$ of about $4.6\times 10^4$, in good agreement with the observed low-temperature $Q$ in our experiment. The corresponding amplitude spectral density of the thermal torque noise, shown in Fig.~\ref{fig:torque_noise}, is about 3.4 times greater than that for a fiber at a uniform temperature of \SI{5}{K} with $Q=4.5\times 10^4$.

\bibliography{CWRSI.bib}

\end{document}